\documentclass{moriond}

\def\be{\begin{equation}}
\def\ee{\end{equation}}
\def\bea{\begin{eqnarray}}
\def\eea{\end{eqnarray}}

\newcommand{\Photo}{}

\begin{document}

\vspace*{4cm}

\title{High resolution SZ observations at the IRAM 30-m telescope with NIKA}

\author{R.~Adam$^1$, 
A.~Adane$^2$
P.~Ade$^3$,
P.~Andr\'e$^4$,
A.~Beelen$^5$,
B.~Belier$^6$,
A.~Beno\^it$^7$,
A.~Bideaud$^3$,
N.~Billot$^8$,
N.~Boudou$^7$,
O.~Bourrion$^1$,
M.~Calvo$^7$,
A.~Catalano$^1$,
G.~Coiffard$^2$,
B.~Comis$^1$,
A.~D'Addabbo$^{7,14}$,
F.-X.~D\'esert$^9$,
S.~Doyle$^3$,
J.~Goupy$^7$,
C.~Kramer$^8$,
S.~Leclercq$^2$,
J.~F.~Mac\'ias-P\'erez$^1$,
J.~Martino$^5$,
P.~Mauskopf$^{3,13}$,
F.~Mayet$^1$,
A.~Monfardini$^7$,
F.~Pajot$^5$,
E.~Pascale$^3$,
L.~Perotto$^1$,
E.~Pointecouteau$^{10,11}$,
N.~Ponthieu$^9$,
V.~Rev\'eret$^4$,
L.~Rodriguez$^4$,
F.~Ruppin$^1$,
G.~Savini$^{12}$,
K.~Schuster$^2$,
A.~Sievers$^8$,
C.~Tucker$^3$,
R.~Zylka$^2$}

\address{$^1$LPSC, Universit\'e Grenoble-Alpes, CNRS/IN2P3, 53, rue des Martyrs, Grenoble, France \\
$^2$Institut de RadioAstronomie Millim\'etrique (IRAM), Grenoble, France \\
$^3$Astronomy Instrumentation Group, University of Cardiff, UK \\
$^4$AIM, CEA/IRFU, CNRS/INSU, Uni. Paris Diderot, CEA-Saclay, 91191 Gif-Sur-Yvette, France \\
$^5$Institut d'Astrophysique Spatiale (IAS), CNRS and Universit\'e Paris Sud, Orsay, France \\
$^6$Institut d'Electronique Fondamentale (IEF), Universit\'e Paris Sud, Orsay, France \\
$^7$Institut N\'eel, CNRS and Universit\'e de Grenoble, France \\
$^8$Institut de RadioAstronomie Millim\'etrique (IRAM), Granada, Spain \\
$^9$IPAG, CNRS and Universit\'e de Grenoble, France \\
$^{10}$Universit\'e de Toulouse, UPS-OMP, IRAP, Toulouse, France \\
$^{11}$CNRS, IRAP, 9 Av. colonel Roche, BP 44346, F-31028 Toulouse cedex 4, France \\
$^{12}$UCL, Department of Physics and Astronomy, Gower Street, London WC1E 6BT, UK \\
$^{13}$School of Earth and Space Exploration and Dep. of Physics, Arizona State Uni., Tempe, AZ 85287 \\
$^{14}$Dipartimento di Fisica, Sapienza Universit\`a di Roma, Piazzale Aldo Moro 5, I-00185 Roma, Italy}

\maketitle\abstracts{High resolution observations of the thermal Sunyaev-Zel'dovich (tSZ) effect are necessary to allow the use of clusters of galaxies as a probe for large scale structures at high redshifts. With its high resolution and dual-band capability at millimeter wavelengths, the NIKA camera can play a significant role in this context. NIKA is based on newly developed Kinetic Inductance Detectors (KIDs) and operates at the IRAM 30m telescope, Pico Veleta, Spain. In this paper, we give the status of the NIKA camera, focussing on the KID technology. We then present observations of three galaxy clusters: RX~J1347.5-1145 as a demonstrator of the NIKA capabilities and the recent observations of CL~J1226.9+3332 ($z=0.89$) and MACS~J0717.5+3745 ($z=0.55$). We also discuss prospects for the final NIKA2 camera, which will have a 6.5 arcminute field of view with about 5000 detectors in two bands at 150 and 260 GHz.}

\section{Introduction}\label{sec:introduction}
The observations of clusters of galaxies, from radio to gamma-ray, have proved to be a powerful probe of large scale structure formation.  Clusters are mainly composed of dark matter ($\sim$ 85\% of their total mass) that drives the gravitational processes. Their baryonic component, mostly made of hot ionized gas, {\it i.e.} the intra cluster medium (ICM), is subdominant but responsible for more complex physics that can limit their use for cosmology. Consequently, a detailed characterization of the complex gravitational and non-gravitational processes acting in galaxy clusters is mandatory not only for a deep understanding of their physics, but also in order to properly use them as robust cosmological probes. One way of detecting and studying them is to use the thermal Sunyaev-Zel'dovich (tSZ) effect \cite{sunyaev1972,sunyaev1980}. When traveling trough the ICM, Cosmic Microwave Background (CMB) photons can Compton inverse interact on the energetic electrons of the hot gas. This leads to a characteristic spectral distortion of the CMB where clusters appear as a decrement in the CMB intensity for frequencies below 217 GHz, and as an increment above (see \cite{birkinshaw1999} for a review on the tSZ effect). Recent tSZ observations have been performed by the South Pole Telescope \cite{carlstrom2011}, the Atacama Cosmology Telescope \cite{Kosowsky2003} and the Planck satellite \cite{planck_overview2013,planck_sz_cat2013} leading to large tSZ selected cluster samples. Nevertheless, the relatively poor resolution of these observations (larger than 1 arcminute) does not allow to probe their inner structure, especially for intermediate and high redshift clusters. In this context, the New IRAM KID Arrays (NIKA) will play a significant role in the study of such systems. NIKA is a high angular resolution imaging camera based on recently developed Kinetic Inductance Detectors (KIDs). This camera is the prototype of the future larger camera NIKA2. 
In this proceeding, we present the status of NIKA, show recent tSZ results, and discuss perspectives for NIKA2.

\section{NIKA: a KID based camera at the IRAM 30-m telescope}\label{sec:NIKAat30m}
In order to push the study of galaxy clusters to higher redshifts, tSZ observations require instruments with high sensitivity, high angular resolution and large mapping speed. Since traditional detectors are already photon noise limited, in particular for ground-based observations, such goals require the development of large arrays of detectors. Cryogenics constraints impose to read as many detectors as possible with a single wire to limit the number of dissipative elements. Kinetic Inductance Detectors (KIDs) offer a promising alternative to other detectors since they are intrinsically frequency multiplexed.
KIDs are high quality superconducting resonator. They are made of a capacitor $C$ and an inductance $L = L_g + L_k$ that is the sum of a geometric and a kinetic part. Absorbed photons of high enough energy can break Cooper pairs in the material (aluminum). The induced variation in the density of the charge carriers changes the kinetic part of the inductance that, in turn, leads to a variation of the resonance frequency of the resonator, $f_0 = \frac{1}{2 \pi \sqrt{LC}}$. This frequency shift can be used to monitor the incoming optical power as both quantities are directly proportional \cite{calvo2012}.

The New IRAM KID Arrays (NIKA) is a KID based instrument that operates at the IRAM 30m telescope, Pico Veleta, Spain. It uses Hilbert dual-polarization design LEKID pixels (Lumped Element KID \cite{doyle2008,roesch2012}) read with the dedicated NIKEL electronics \cite{bourrion2011}. Two arrays of 132 and 224 detectors simultaneously image astrophysical sources with beam FWHM of 18 and 12 arcsecond at 150 and 260 GHz respectively. The NIKA sensitivities have improved continuously with development and are now reaching state-of-the-art performance at both frequencies (see \cite{monfardini2010,monfardini2011,calvo2012,catalano2014} for more details).
NIKA is the prototype of the final camera NIKA2, which will cover a 6.5 arcminute instantaneous field of view with 1000 detectors at 150 GHz and 4000 at 260 GHz. This future camera will also have polarization capabilities at 260 GHz.

\section{Sunyaev-Zel'dovich observations with NIKA}\label{sec:SZresult}
\subsection{A dedicated tSZ analysis}\label{sec:SZana}
In order to recover the diffuse tSZ emission, we make use of the dual-band capability of NIKA to separate the atmospheric component and the tSZ one. To do so, we take advantage of their different dependance with the observing frequency $\nu$: the atmospheric noise is roughly proportional to $\nu^2$ while the tSZ signal follows the characteristic spectrum given  by
\begin{equation}
	\frac{\delta I_{\mathrm{tSZ}}}{I_0} = -y \ \frac{x^4 e^x}{\left(e^x-1\right)^2} \left(4 - x  \ \mathrm{coth}\left(\frac{x}{2}\right) \right),
	\label{eq:sz}
\end{equation}
where $x = \frac{h \nu}{k_B T_{\rm CMB}}$ is the dimensionless frequency, $y = \frac{\sigma_{\mathrm{T}}}{m_{\mathrm{e}} c^2} \int P_{\mathrm{e}} dl$ is the Compton parameter (that gives the amplitude of the tSZ effect integrating the electronic pressure $P_e$ along the line-of-sigh), $T_{\rm CMB}$ is the CMB temperature. For more details concerning the tSZ mapmaking process, see \cite{adam2013}.

\subsection{RX~J1347.5-1145}\label{sec:RXJ1347}
\begin{figure*}[ht]
\centering
\includegraphics[width=5.2cm]{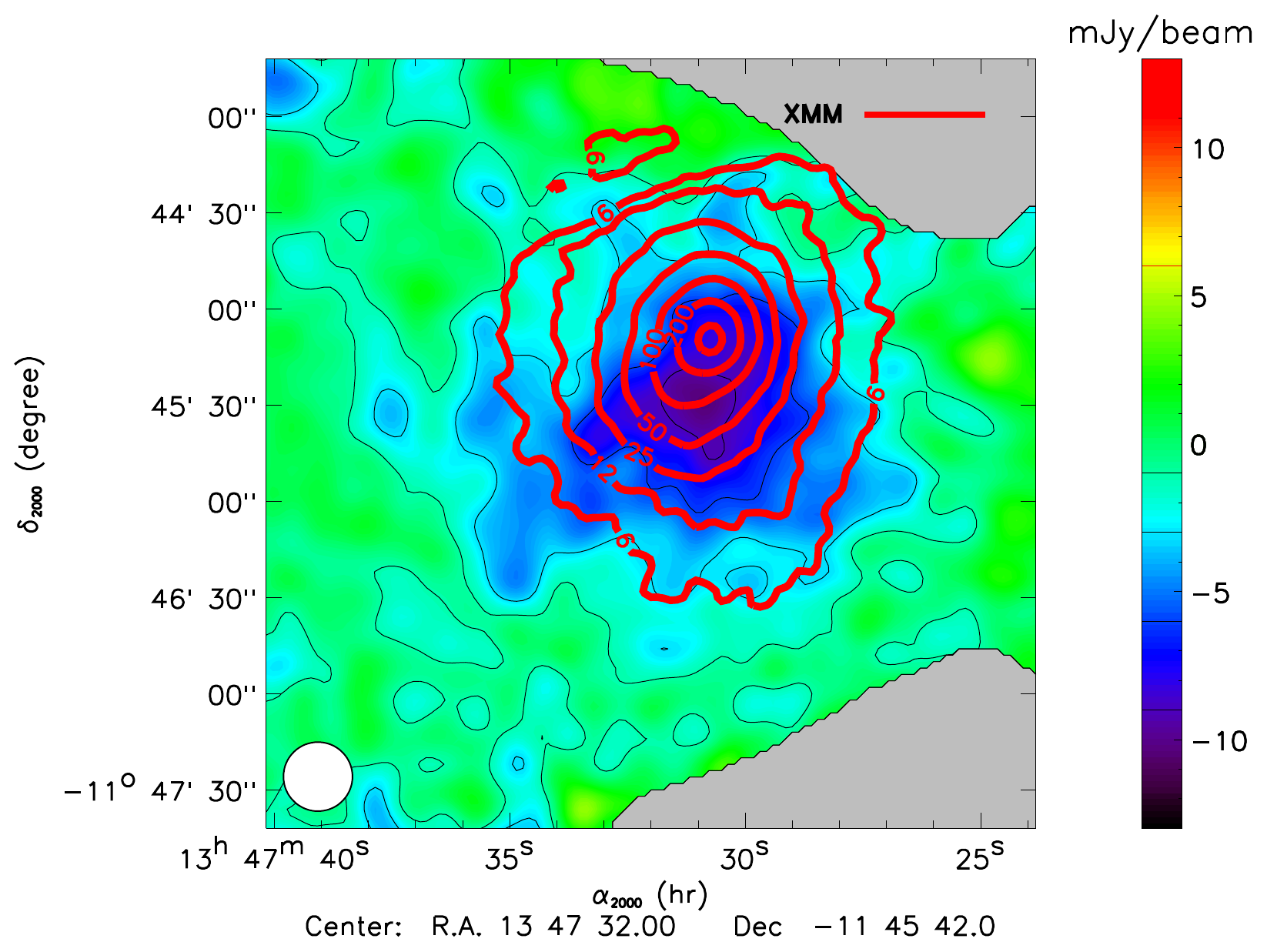}	
\includegraphics[width=5.2cm]{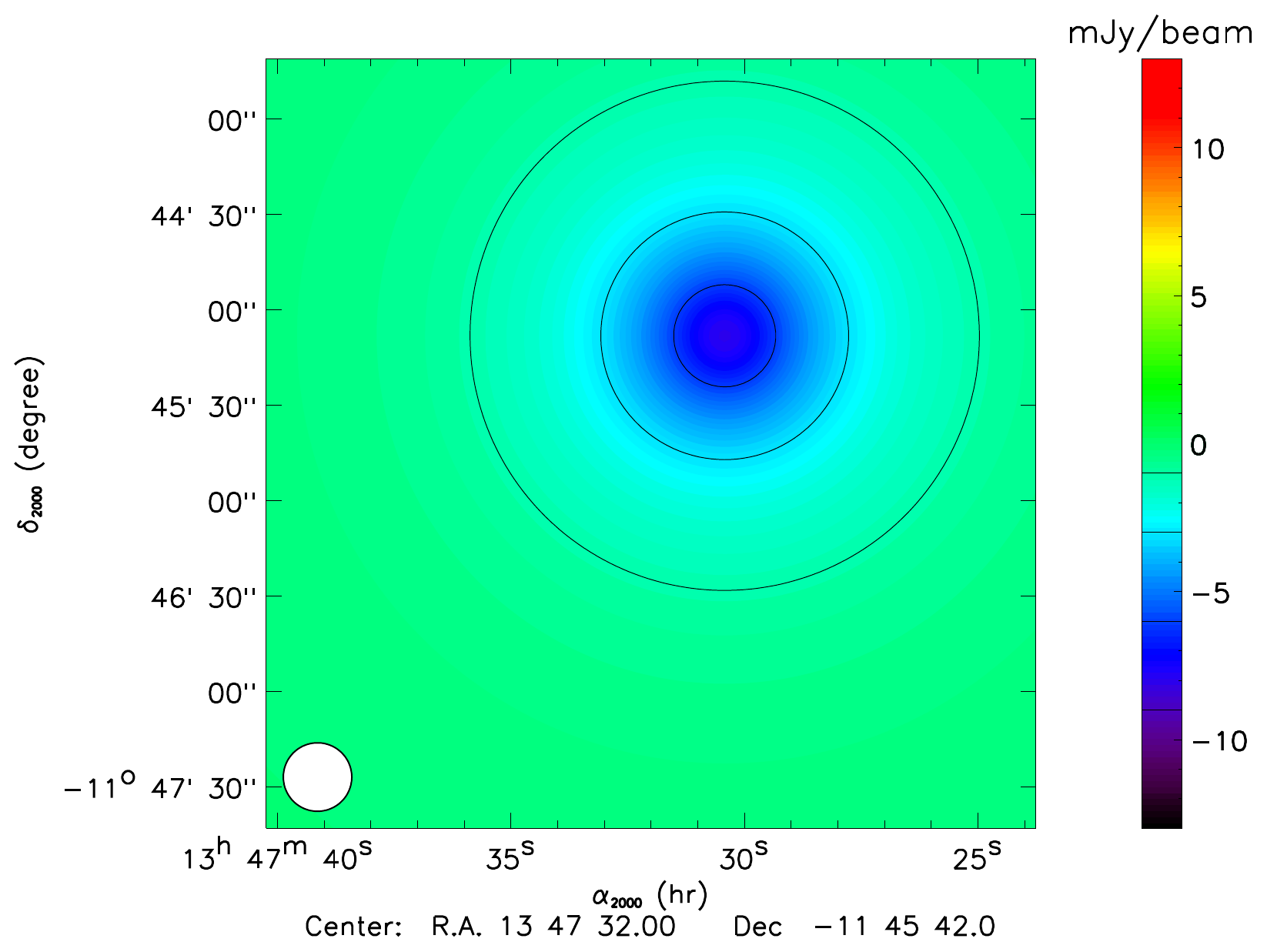}
\includegraphics[width=5.2cm]{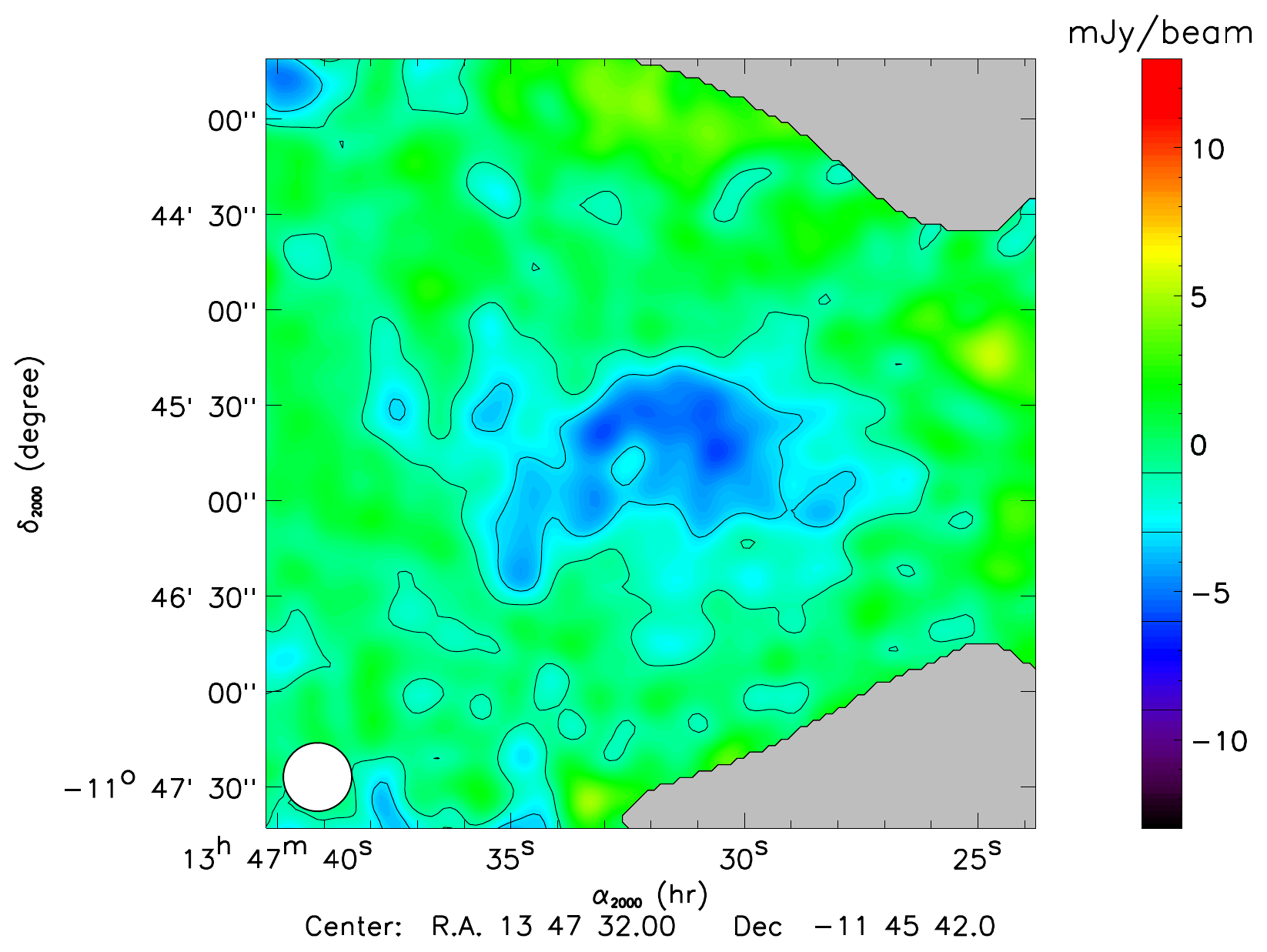}
\caption{Left: 140 GHz flux density map of the intermediate redshift cluster ($z=0.45$) RX~J1347.5-1145. A $\sim$ 4 mJy radio source is located within 3 arcseconds of the X-ray center, and has been remove in these maps. XMM X-ray photon counts contours are over-plotted in red. Middle: best fit model map of the northern part of the cluster. Right: residual map highlighting the overpressure caused by the merger on the south-east of RX~J1347.5-1145. The maps have been smoothed with an 11 arcsecond gaussian filter.}
\label{fig:RXJ1347}
\end{figure*}

The cluster RX~J1347.5-1145 was observed in November 2012 as a demonstrator of the tSZ capability of a KID based instrument. The left panel of Fig.~\ref{fig:RXJ1347} shows the map at 140 GHz. In the South-East, a sub-cluster is merging with the main cluster (see for example \cite{plagge2012,allen2002} and references therein). This heats the ICM and increases the tSZ signal at about 20 arcsecond on the SE of the X-ray peak. X-ray contours from XMM photon count are displayed on the map and highlight the merger. As X-ray emission is proportional to the density squared and depends weakly on the temperature, we can see the expected tSZ extension on the NIKA map. In order to model the structure of the cluster, we fit a generalized NFW pressure profile model \cite{nagai2007} centered on the X-ray peak, excluding the southern shocked area and subtract it from the whole map. The best fit model and the residual are given respectively on the middle and right panel of Fig.~\ref{fig:RXJ1347}. Details on the observation and analysis can be found in \cite{adam2013}.

\subsection{Observations of February 2014}\label{sec:Feb2014}
\begin{figure*}[ht]
\centering
\includegraphics[width=5.2cm]{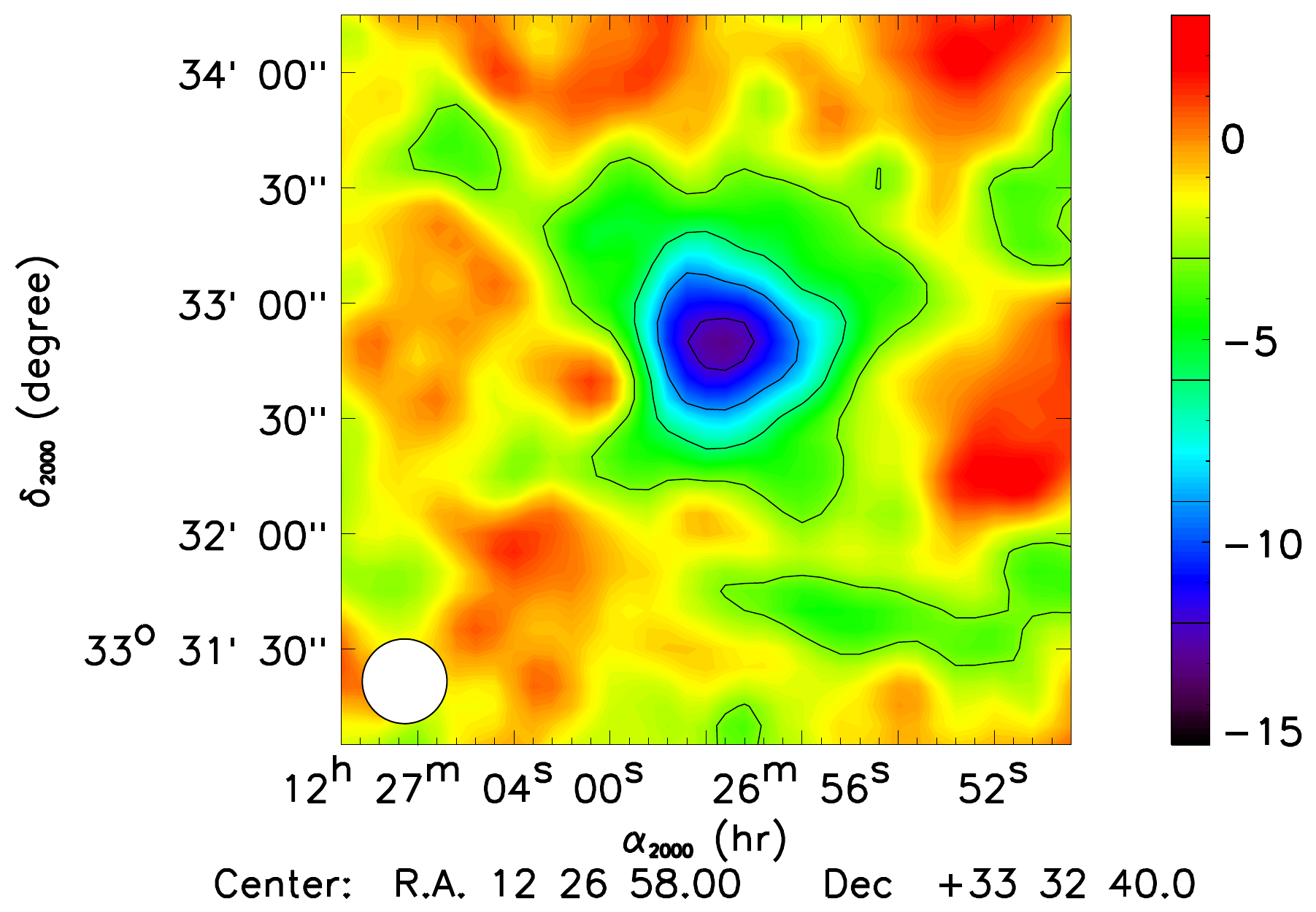}
\includegraphics[width=5.2cm]{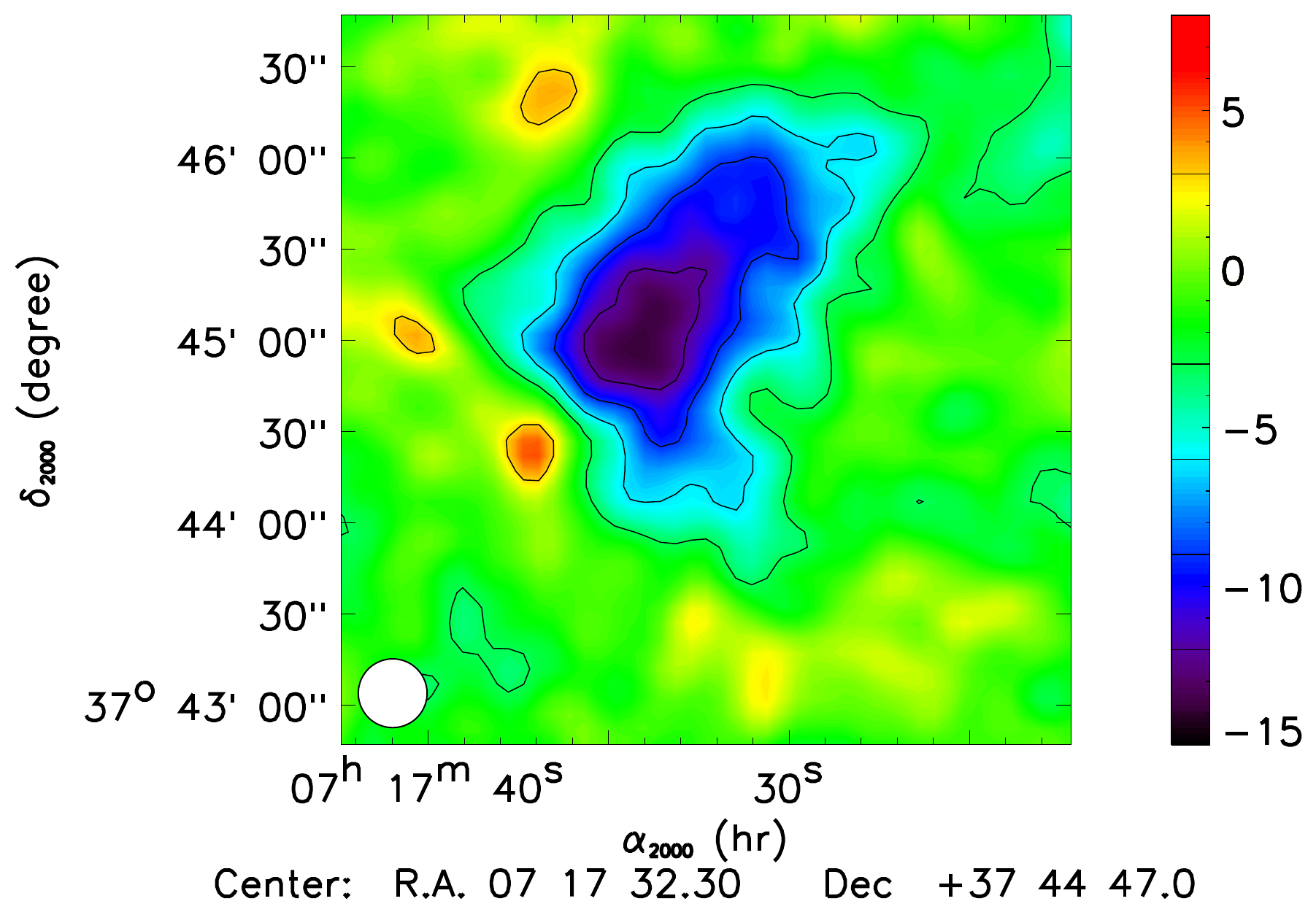}
\caption{150 GHz signal to noise map of CL~J1226.9+3332 (left) and MACS~J0717.5+3745 (right). The clusters are clearly detected with a peak tSZ decrement reaching a SNR of about 15. The contours are 3, -3, -6, -9 and -12 $\sigma$. The maps presents diffuse emission up to more than 1 arcminute radius. The effective beam is shown on the bottom left part of the maps and accounts for the 13 arcsecond gaussian smoothing and the 18 arcsecond native NIKA beam, {\it i.e.} a total of 22 arcsecond. Note the presence of a foreground galaxy detected on the south-east of MACS~J0717.5+3745.}
\label{fig:CL1227}
\end{figure*}

The two clusters CL~J1226.9+3332 and MACS~J0717.5+3745, at $z=0.89$ and $z=0.55$ respectively, have been observed during the first NIKA open pool in February 2014. In both cases, we detect the clusters with signal to noise of about 15 at 150 GHz. The two maps present diffuse emission and will be used to study the ICM and the dynamical state of the clusters in a forthcoming paper. Here, we only show their signal to noise maps in Fig.~\ref{fig:CL1227}.

\section{Conclusions and perspectives}\label{sec:conclusion}
The use of galaxy clusters as a cosmological probe will require high resolution tSZ observations. The NIKA camera has proved to be an instrument capable of such study with the observation of RX~J1347.5-1145 at $z=0.45$. Using a gNFW pressure profile model of the cluster, we have further characterized the cluster ICM and highlighted the merging of a sub-cluster on its south-east part. We also report observations of the clusters MACS~J0717.5+3745 and CL~J1226.9+3332 that are respectively a dynamically complex system and a high redshift cluster (z=0.89). The NIKA2 camera is being built and will replace NIKA in 2015. It will push tSZ studies a step further. It will enable the characterization of the tSZ morphology of clusters at both large and small angular scales. NIKA2 will be used to calibrate the tSZ flux as a mass proxy and its evolution with redshift precisely. It will also help to characterize the dynamical state of clusters of galaxies.

\section*{Acknowledgments}
{\footnotesize We would like to thank the IRAM staff for their support during the observations. 
This work has been partially funded by the Foundation Nanoscience Grenoble, the ANR under the contracts "MKIDS" and "NIKA". 
This work has been partially supported by the LabEx FOCUS ANR-11-LABX-0013. 
This work has benefited from the support of the European Research Council Advanced Grant ORISTARS under the European Union's Seventh Framework Programme (Grant Agreement no. 291294).
The NIKA dilution cryostat has been designed and built at the Institut N\'eel. In particular, we acknowledge the crucial contribution of the Cryogenics Group, and in particular Gregory Garde, Henri Rodenas, Jean Paul Leggeri, Philippe Camus. 
R. A. would like to thank the ENIGMASS French LabEx for funding this work. 
B. C. acknowledges support from the CNES post-doctoral fellowship program. 
E. P. acknowledges the support of grant ANR-11-BS56-015.
We would like to thank C\'eline Combet for useful comments.}


\end{document}